# A Habitable Zone Census via Transit Timing and the Imperative for Continuing to Observe the Kepler Field

**Authors:** Daniel C. Fabrycky (U Chicago), Eric B. Ford (Penn State U), Matthew J. Payne (Harvard-Smithsonian Center for Astrophysics), Jason Steffen (Northwestern U), Darin Ragozzine (Florida Institute of Technology), Tsevi Mazeh (Tel Aviv U), Jack J. Lissauer (NASA Ames), William Welsh (San Diego State U)

## Abstract
We propose a scientific program to complete a census of planets, characterizing their masses, orbital properties, and dynamical histories using continued observations of the *Kepler* field of view with the *Kepler* spacecraft in a two reaction wheel mission (subsequently referred to as *Kepler-II*). Even with a significantly reduced photometric precision, extending time-domain observations of this field is uniquely capable of pursuing several critical science goals: 1) measuring the architectures of planetary systems by identifying non-transiting planets interleaved among known transiting planets, 2) establishing the mass-radius relationship for planets in the important transition region between small, gas-rich sub-Neptune planets and large, rocky super-Earths, and 3) uncovering dynamical evidence of the formation and evolution of the inner regions of planetary systems. To meet these objectives, the unique multi-object observing capabilities of *Kepler* will be used in a set of concurrent campaigns with specific motivations. These campaigns focus largely on the ability to interpret Transit Timing Variations (TTVs) that result from dynamical interactions among planets in a system and include: 1) observations of systems that exhibit large TTVs and are particularly rich in dynamical information, 2) observations of systems where additional transit times will yield mass measurements of the constituent planets, 3) observations of systems where the TTV signal evolves over very long timescales, and 4) observations of systems with long-period planet candidates where additional transits will remove orbital period ambiguities caused by gaps in the original *Kepler* data.

## Introduction

With the *Kepler* spacecraft's reduced pointing capabilities, a number of options are being considered for how best to utilize its remaining, high-quality photometric capabilities. Here we present a number of science questions that can be uniquely addressed with continued *Kepler* operations in the original *Kepler* field of view. One of the primary advantages that continued observations of the *Kepler* field of view offers is its ability to leverage the existing 4-year observation baseline (an exceptionally valuable dataset) and expand it in the direction that presents the greatest challenge for any existing or future study of exoplanets, but one that the *Kepler* spacecraft is uniquely capable of providing – time.

Even with a diminished capacity, Kepler is still capable of answering key scientific questions that otherwise may be left unanswered for decades. Of particular importance are the properties of planets and planetary systems in the region from a few stellar radii to just beyond that star's habitable zone. Some of the ground-breaking discoveries that *Kepler* has made have raised important questions that pertain directly to the habitability of planets, how they form and evolve dynamically, and how common they are. Many of these questions are best addressed with long-duration photometric time series for those systems which enables more detailed dynamical analysis.

We give three primary scientific objectives for this proposed work. These objectives will be met through a combination of additional *Kepler* data and subsequent analysis. We expect a reduction of the number of observable targets in *Kepler-II*, relative to the original survey. The target selection for *Kepler-II* will be guided by the needs of a number of observation "campaigns" that are outlined below.

1) **Constructing full portraits of planetary systems, based on transits alone:**
The transit timing of multiply-transiting planetary systems can yield exquisite information on the observed planets. However, geometry dictates that a transiting configuration exists only for a narrow range in inclination to the observer. Inevitably, additional non-transiting planets reside in many of these systems. Since most of the *Kepler* target stars are too faint for intensive high-precision radial velocity (RV) campaigns, these additional planets are better studied through transit timing variations (TTVs) – the deviations from a constant orbital period due to planet-planet gravitational interactions. Planet detection was, actually, one of the original motivations for transit timing studies (Holman & Murray 2005, Agol et al. 2005). To date, three non-transiting planets have been found by applying this technique to *Kepler* data (Ballard et al. 2011; Nesvorny 2012, 2013). All three of these systems consisted of a single transiting planet and a single non-transiting planet. With additional data in the *Kepler* field we can will expand the use of this technique by applying it to systems that already have multiple, transiting planets. This analysis will enable a full characterization of these systems, which is currently unachievable by any other means. Even non-detections provide valuable information about the absence of low-mass planets near resonances with the planets observed to transit.

2) **Surveying planet masses in the habitable zone:**
The large population of sub-Neptune planets was a surprising discovery from the *Kepler* mission. Many questions remain about the formation and masses of planets whose sizes lie between that of the Earth and that of Neptune – a critical transition region as it pertains to habitability. If this type of planet is common in the habitable zone (just as they are common interior to that zone), then it casts doubt on the habitability of radial velocity detections of many known "super-Earths" (m sin i < 10 M_Earth). Mass measurements of these planets is an important ingredient to understanding their potential habitability and the frequency of rocky planets in the habitable zone. TTVs are the best available method for making these measurements. Indeed, TTVs have been the primary means of mass measurement for sub-Neptunes identified by Kepler. An extended time baseline is needed to measure masses via TTV, in and around the habitable zones of *Kepler* targets.

3) **Probing the formation and evolution of rocky planets:**
In contrast to the hot-Jupiters, for which there is evidence of a dynamically violent past (e.g., Winn et al. 2010, Steffen et al. 2012), intermediate-period gas giants in systems with several transiting planets show low eccentricities and inclinations (Albrecht et al. 2013). However, these orbital periods are still too small for the planets to have plausibly formed in-situ; moreover the common presence of planets near mean-motion resonances (MMRs) point to a history of quiescent orbital migration. These systems provide the clearest evidence for large scale disk migration; studying them further will constrain their resonant interactions, yielding insight into their migration history and the physics of their late-stage protoplanetary disks. In contrast, super-Earth and sub-Neptune planets, even those near MMR, may form in-situ and undergo little or no migration – not needing, for example, to cross the system's ice line. Identifying the distinguishing characteristics of systems that undergo large-scale orbital migration will tell us much about the formation and survival of rocky planets in the inner parts of planetary systems. Access to this information requires more Kepler data to characterize the systems.

## Basics of TTVs

The identification of stars with multiple transiting planets is one of the greatest discoveries of the *Kepler* mission. Now, more than 400 of these "multi-transiting systems" containing more than 1000 high probability candidate planets (Burke et al. 2013, in prep.; Lissauer et al. 2012), these multi-transiting systems are the most information-rich planetary systems outside our own solar system (Ragozzine & Holman 2010). This richness comes because they combine the detailed planetary and orbital information available from transiting planets (Winn 2010) with the ability to apply multi-planet orbital dynamics (Lissauer et al. 2011, Fabrycky et al. 2012b). For simplicity, we will refer to these planet candidates as "planets" throughout this manuscript, recognizing both that they largely remain planet candidates, and yet the probability that they are true planets is very high (≥ 98%, Lissauer et al. 2012).

Transit Timing Variations (TTVs) are deviations of the transit epochs from a linear ephemeris due to gravitational planet-planet interactions (Agol et al. 2005, Holman & Murray 2005). The amplitude of the TTV signal scales both with planet mass and with orbital period – allowing the masses of even small, rocky planets to be measured. A TTV signal is also characterized by the timescale of its variations. The closer a planet pair is to mean-motion resonance (MMR) the longer the time required to complete a full TTV cycle (and generally the larger the amplitude of the TTV signal). This fact implies that the longest TTV signals are particularly interesting because proximity to MMR is a strong indicator of the formation and dynamical evolution of the system and because planets with lower masses are able to produce detectable TTV signals on longer timescales. To date, Kepler TTVs have been the most successful method both for dynamically confirming the planet nature of *Kepler* candidates and for measuring the mass for small planets (e.g., Lissauer et al. 2011a, Fabrycky et al. 2012a, Steffen et al 2012a, Ford et al. 2012a, Xie 2012, Wu & Lithwick 2012).

Gleaning mass measurements from the TTV signal is non-trivial, strongly dependent on several orbital parameters (e.g., period, eccentricity, pericenter). Distinguishing among possible solutions for a given system requires sufficient data on that system. TTVs have been detected in a large number of Kepler systems (e.g., Ford et al. 2011, 2012b; Mazeh et al. 2013), and the distinguishing features of those signals are, in many cases, only beginning to appear. This fact motivates continued observations of these systems since the value of additional data on known systems is much larger than new data on new systems.

Even though increased pointing error will impact *Kepler's* photometric precision, the measurement of the mid-transit time using a known transit shape (known from prior Kepler data and modeling) is quite robust to systematic errors – even on the timescale of a single transit. Thus, transit timing measurements are expected to be quite good even with pointing errors.

By extending TTVs measurements beyond the current time series, *Kepler* can significantly increase our ability to measure the masses of large numbers of small planets. This directly addresses one of *Kepler's* the most critical questions: *What is the mass-radius relation of small planets?* Recent Kepler mass measurements indicate that some planets with radii less than twice that of Earth – the so-called "super-Earths" – are actually low density "sub-Neptunes" (e.g., Kepler-11 planets, Lissauer et al. 2011a, Rogers et al. 2011). Sub-Neptunes are poor candidates for potentially habitable planets, despite their roughly Earth-like size, because of their large gaseous atmospheres. Thus, a primary astrobiological science driver, the frequency of "Earth-like" planets (η-Earth), requires measuring planet masses as well as sizes, so as to recognize which small planets are rock or water and which are primarily gaseous. Measuring the frequency of potentially habitable rocky planets requires knowledge of the mass-radius relationship of small planets, those in the transition from gas rich sub-Neptunes to rocky super-Earths. This knowledge is best obtained through long-baseline TTV measurements – which *Kepler* is unique in its ability to provide.

Most TTV signals have timescales much longer than the 4-year primary mission. In this regime, new data added to the end of the time series for a system has much higher impact than simply starting a time series on a new discovery. If the photometric precision remains constant, the information in a TTV time series in the current regime scales as $t^{5/2}$. If the timing precision of observations during the next 4 years is 3, 10, or 30 times worse than during the first 4 years, then the TTV information content increases by a factor of 3.16, 1.6, or 1.07, respectively.

The only instrument capable of making these valuable transit measurements on any sizeable number of *Kepler* targets is *Kepler* itself. The number of apertures may be reduced significantly from the main mission, but other telescopes – HST, Spitzer, SOFIA, and ground-based telescopes – are generally limited to *one* target at a time, and only over-subscribed resources are able to reach the precision of *Kepler-II*. In theory, one could continue monitoring up to about 2 dozen TTV targets, but with *unrealistic* work, cost, and coordination. No follow-up effort on the scale we propose could be done without the *Kepler* spacecraft itself.

## Kepler-II primary campaigns

Here we outline several science campaigns that motivate target selection, allowing the scientific objectives of the introduction to be realized. These primary campaigns are: 1) Planets exhibiting large amplitude TTV signals, 2) Planet pairs with sinusoidal TTV signals, 3) Planets with very long-duration TTV periods, and 4) Long-period planets. We discuss each of these campaigns in turn.

### Planets with High Signal-to-Noise TTV Signals

A significant number of Kepler planet candidates have amplitudes so large that they are capable of providing exceptionally high precision measurements of masses and orbital elements. The first published example of a system like this is Kepler-9 (Holman et al. 2010), which now has a completely defined dynamical model with parameters (e.g., planetary masses, eccentricities, inclinations) measured to a few percent. Another system, Kepler-30 (Fabrycky et al. 2012a, Sanchis-Ojeda et al. 2012), has three planets, and similar constraints. In Kepler-9 and Kepler-30, the interacting planets all have transit signals. Three other systems that have only a single transiting planet, but with TTV amplitudes of several hours yield, providing unique orbital solutions: KOI-872 (Kepler-46), KOI-142, and KOI-1474 (Nesvorny et al. 2012,2013, Dawson et al. 2012).

Although these systems have well-constrained dynamical solutions, all such models would be compromised if additional planets cause TTV on a longer timescale. Thus, continued long-term monitoring by *Kepler* would either (a) discover additional unseen planets, completing the survey of the architectures of these systems, or (b) provide more stringent limits on additional planets, thereby increasing the robustness of mass and density measurements of transiting planets from TTVs. In either case, *Kepler's* value for characterizing such systems is unique.

There are a few dozen more systems with TTV amplitudes above 1 hour, whose transit timing signature could be easily measured even with *Kepler*'s reduced photometric precision. These systems have not been published at this time simply because the TTV model does not yet uniquely identify the perturber. A longer baseline would resolve remaining degeneracies in many systems, as the predicted timings of the models diverge in the future (see Figure 1). After the qualitative ambiguities of the solutions are solved, these systems can be used for fundamental planetary astrophysics, i.e. measurements of orbital architectures and new points for the mass/radius diagram.  Another type of system in need of better solution is transiting circumbinary planets, of which 13 have been detected.  Many of them have fewer transits measured than free parameters; thus, their orbital solutions will improve dramatically with additional transit observations.

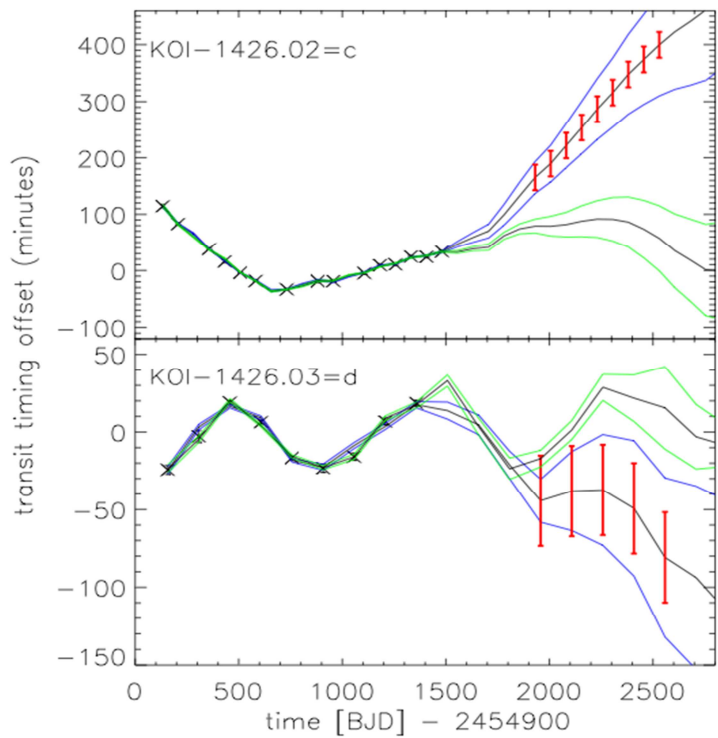

*Figure 1.*  Transit timing observations (x's, with very small error bars) and predictions for KOI-1426.  Two other planets are included in the modeling, but are not represented in the figure: an inner transiting planet (KOI-1426.01=b) with well-modeled timing variations, and two different possibilities for an external, non-transiting planet (e) with a period of either 258 days (blue) or 352 days (green).  Kepler-II observations (red bars, with the scale of timing errors corresponding to an instrumental noise of 1000 ppm in 6-hour intervals) will be able to rule out one of these models, then provide tight constraints on orbital parameters, masses, and radii.

Of particular importance, some systems have large amplitude TTV signals and such strong perturbations that there is a very real possibility of losing the ephemeris (i.e., no longer knowing whether a future transit observation corresponds to the Nth or N+1th transits) if *Kepler* observations cease now – making reliable ground-based transit observations impossible to schedule.  For roughly a dozen systems, more *Kepler* data *must* be obtained if we are to continue any meaningful long-term dynamical studies of these systems which are, as indicated by their large TTV signals, exceptionally rich in information about planet masses, densities and their dynamical history.  Therefore, *Kepler*'s unblinking eye is a necessary asset.

### Planet Pairs with Sinusoidal TTV Signal

For the majority of TTV signals seen to date in the data, the transit times can be adequately represented by a linear ephemeris plus a sinusoidal fluctuation.  If the planet causing these fluctuations also transits, it tells us the forcing function that one planet exerts on the other, and from there we can from there precisely calculate the period of the TTV signal, $P_{TTV}$.  From this information, analytic formulae exist to interpret the amplitudes and relative phases of the TTV signals from each planet, as a function of the planet masses and orbital eccentricities (Lithwick et al. 2012).

For some systems where multiple planets are seen, the TTV period is predicted to be so long that the current data is not sufficient determine the TTV phases and amplitudes.  *Kepler-II* can make these measurements for the planet pairs where the transits are detected at >3 sigma (or potentially even lower SNR), and where the TTV periods are between 1067 days and 1687 days (we assume that 1.5 TTV periods need to be observed to have confidence in the measurement, and that *Kepler-II* will start soon and run for two years).

Using the KOIs on NExScI's list (8/30/13) that have a disposition of "candidate," we identified planet pairs that lie near resonances and that have an expected $P_{TTV}$ in the above range. If *Kepler-II* is 5 times (2 times) less precise than *Kepler* during its main mission, then we will profitably observe 7 (19) systems and measure their long-timescale TTV curves. Thus of order ten more multi-transiting planetary systems can be dynamically characterized after *Kepler-II*, which are not possible with *Kepler* data alone and which would be exceedingly difficult to characterize by any other means. These systems will provide mass measurements of their planets, giving important constraints on the composition of the planets – distinguishing between rocky super-Earth planets and gaseous sub-Neptunes.

## Planets with Long TTV Periods

As shown by many theoretical studies (e.g., Goldreich and Schlichting 2013, and references therein), the capture of a pair of planets near a mean motion resonance (MMR) depends on a few factors, including the mass ratio of the two planets relative to the central star, the mass in the disk, and the processes driving the migration. In particular, the final state of the pair, and the proximity of the two periods to their resonance, depends on the details of the interaction between the two planets and the disk. Therefore, the population of pairs of planets near MMR (e.g., Fabrycky et al. 2012b) can serve as the Rosetta stone for the understanding the planetary migration, a key element in our understanding of planet formation. In particular, we are interested in how close the pairs of planet are to resonance.

The population of Kepler pairs of planets studied so far is based only on pairs of *transiting* planets. For this, the two planets nearly always have very small mutual inclinations. Those studies are not able to probe the population of pairs of planets that have even modest mutual inclinations, where only one planet is transiting its parent star. For this reason, the known pairs of transiting planets discovered by Kepler might represent only the tip of the iceberg for planets near MMR. Moreover, this tip is biased toward interactions that yield highly coplanar systems – implying dynamical interactions with a disk and perhaps only those with certain disk properties.

Fortunately, TTVs provide a way to study pairs of planets near MMR with large relative inclination, if just one of the planets is a transiting planet. The period of the TTV modulation ($P_{TTV}$) is inversely proportional to the proximity of the pair to their resonance. Thus, for pairs that reside very close to their resonance, hence those systems with large signals, $P_{TTV}$ can be quite long, of the order of several hundreds to a few thousands days. In order to study this population of planet pairs and uniquely characterize the non-transiting member, we need a baseline of observations longer than $P_{TTV}$ itself.

Systems with TTV periods shorter than 1000 days have been used to derive the mass and period of the non-transiting planet and to estimate how close the pair is to the resonance (e.g., Nesvorny 2012). However, in the present Kepler data we have detected **50** planets with a TTV period longer or equal to ~1000 days and with amplitudes of the order of hundreds of minutes (Mazeh et al. 2013). Such large-amplitude TTV signals can be induced only by an unseen planet near MMR. Unfortunately, the present time baseline is not sufficient to measure the TTV period and more data are needed. Extending *Kepler* observations of these targets will enable us to complete this study.

## Long-Period Planets

Another important scientific result that *Kepler* can uniquely provide is the correct measurement of the orbital periods for a number of long-period, transiting planet candidates, completing a census of planets beyond the habitable zone. Such discoveries will reap simple, direct benefits from an extended observational baseline including the following.

1) Correct derivation of the planet orbital period. There are cases among the current candidates in which the low number of transits (e.g. N=1 or 2), combined with gaps in the observational data during Q 1-16, yields ambiguous orbital periods. Examples include KOI 1463.01 and 375.01 (Borucki et al. 2011, Batalha et al. 2013), for which both have a data gap mid-way between the two known transits leading to a period ambiguity of ~500 days versus ~1,000 days for both systems. Establishing correct orbital periods in this way is extremely *in*sensitive to photometric precision: *any* additional transit signal is sufficient to confirm or correct the orbital period. Many of these systems have too large of an ephemeris uncertainty to reasonably attempt a ground-based campaign of sufficient duration. Exploiting Kepler's ability to monitor large numbers of planets for long periods is the best way to recover these rare systems that provide information about the atmospheres of cool giant planets and the architectures of planetary systems with such planets.

2) Improved measurement of transit properties. While of lower precision, additional transit data will improve our knowledge of the period on which to fold the data, thus allowing the planetary and orbital parameters to be more accurately determined from the folded transit light curves. Simply obtaining 1 or 2 more transits for many long period systems will significantly decrease the number of systems with false periods, as well as significantly improving the measured properties of their transits. Figure 2 illustrates the 51 long-period planets with N<4 known transits (Huang et al 2013, Wang et al. 2013, Batalha et al. 2013), plotting period against transit depth. Using the horizontal line at 2,000 ppm as a guide, 34 (48) systems will have improved period measurement and light-curve fitting as a result of a 2-year (4-year) *Kepler-II* mission. This includes 8 (11) long period systems that will be confirmed (the number of measured transits is increased to greater than 2) during such a 2-year (4-year) mission.

The science accessible from measured transit times of long period (~ 1-4 years) planets are distinct from those of short period planets. For these planets, although there will not be sufficient transits to fully characterize the systems, the presence or absence of TTVs and a measurement of their magnitude gives insights into the perturbations to which the planet is subjected, and therefore to the dynamical activity of the system and the overall system architecture. *Kepler* has already told us a great deal about the abundance of planets and architectures of planetary systems, but that information is primarily for planets with periods of less than ~6 months. With extended observations of planets with orbital periods of order one year, we will gain insight into the presence of perturbing planets within and beyond the habitable zones of solar-type stars.

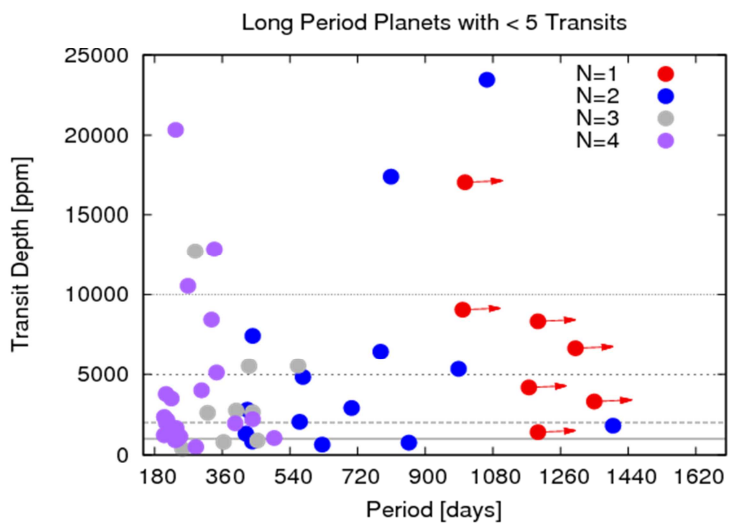

**Figure 2.** *The periods and transit depths for the current long period planets with <5 transits. The periods for the N=1 systems are lower bounds estimated using data gap from the observed transit to the end of the extant data (Q1-16). The horizontal lines at 1, 2, 5 & 10 thousand ppm provide a conservative indication of the number of long period systems for which improved precision will be gained during Kepler-II.*

So far, fewer long-period planets exhibit TTVs than short period planets do, simply due to fewer transit observations. However, the sensitivity of TTV measurements of long-period planets to perturbations of outer planets is very good since the TTV signal scales linearly with the orbital period. These larger TTV signals compensate for the smaller number of transits observed. A minimum of three observed transits is required to detect TTVs, but the addition of a fourth or fifth transit are extremely useful as three transits can happen to fall periodically even in the case of large-amplitude TTV.

Characterizing these long-period planets is extremely valuable for our understanding of planets in the habitable zone, since they tell us the kinds of planets that typically live beyond that zone. Measuring dynamical perturbations will connect the architectures of the inner parts of planetary systems to the outer parts (e.g., the connection between the solar system's terrestrial planets and its giants). They will also provide information about "Jupiter analogues" that is complementary to data from RV surveys.

## Technical Feasibility

*How the Focal Plane will be Used:* Targets, with the number of targets depending on realized photometric precision.

*Class of Science Target:* Point sources, existing KOIs (plus some existing *Kepler* targets from PlanetHunters.org). See supplement at http://www.personal.psu.edu/ebf11/data/kepler/ for target names.

*Number of Targets:* We recommend ~150 targets to survey as the most valuable targets for TTV science. Of course, the actual number will depend on the realized photometric precision: see figure 3. If it were practical to continue observing all KOIs, this would significantly improve searches for TTVs in existing systems and searches for giant planets at larger orbital periods.

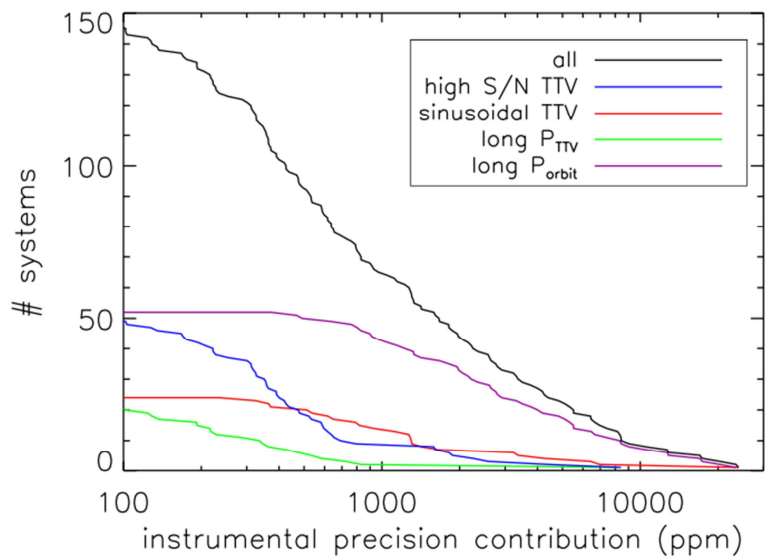

**Figure 3.** *Number of targets. Each of the campaigns is able to reach a certain number of targets as a function of the instrumental precision, which we assume adds in quadrature with the main-mission's 6-hour combined differential photometric precision (CDPP). We assume the high S/N TTV project requires total Kepler-II transit S/N=20, the sinusoidal and long-period TTV projects require total Kepler-II transit S/N=3, and the long orbital period project requires the noise to be less than the depth.*

*Suggested Integration Times:* For the calculations above, we assume a nominal 30-min long cadence observation. The nominal 30-min cadence is not ideal for TTVs of small planets; flux measurements every 2-3 minutes in order to resolve the ingress/egress of these small planets improves the transit timing precision by a factor of ~2, even in the case of white noise (Carter & Agol, in prep.). Shorter cadence would also mitigate the increased systematic errors expected from *Kepler-II*. However, a shorter cadence significantly increases the required data rate. Depending on the number of pixels per aperture required for the adopted observing strategy, there may be a minimum integration cadence or a maximum number of targets. Therefore, we recommend choosing the integration time to be as short as possible while still observing the number of targets for the science goals, given practical constraints such as aperture sizes and frequency of data downlinks to Earth. Note that we do *not* require photometry to be stable on timescales longer than a few transit durations (several hours to two days depending on specific planet). Thus, the photometric requirements are significantly relaxed compared to a planet *search* program.

*Expected Data Storage Needs:* We propose a *Kepler-II* mission focused on follow-up, rather than a planet search. Since the number of targets would be significantly reduced relative to the prime mission, we expect it would be possible to observe all of our targets with significantly larger apertures and a shorter cadence. While TTV targets identified here are the most valuable for additional follow-up, another trade-off that will need to be considered is the opportunity to continue monitoring all KOIs and EBs. In any case, we would recommend choosing the number of targets and integration time so as to use make use of all available storage between data downlinks.

*How Long the Science Program Should be Run:* Since the information content of TTV observations increases faster than linearly with increasing time span, we recommend that the science program continue as long as possible. We note that there are specific times when the scientific value of observing with certain targets is predictably much higher than the average *Kepler* observation. For this reason, one possibility for *Kepler-II* is a mode focused on observing the most interesting targets at the most interesting times. In our recommended baseline *Kepler-II* mission, we could observe specific targets near the times of transits (accounting for the unknown TTVs). One could even choose which spacecraft orientations and CCD modules are used for the observations to improve the precision for those targets. This flexibility allows us to work within wide engineering constraints.

If the primary goal of *Kepler-II* is long baseline studies, it may be worth considering (from a scientific and engineering standpoint) whether taking significant resting periods could allow *Kepler* to continue in such a minimal observing mode for many years. For example, this might be an optimal strategy if obtaining high precision photometry requires so much data be downloaded that the limiting resource becomes DSN time.

*Potential Coordination with Additional Science Programs:* In the event that the *Kepler-II* mission consists of multiple science goals, it is important to remember that almost all of our proposed science do not have specific timing requirements. For example, our science case gains strength during the time it will take to iron out engineering and technical details of the new observing model. Similarly, we note the possibility of meeting multiple science goals by interleaving observations of the *Kepler* field with observations of another field for other purposes. While any time away from continuous observations of the *Kepler* field certainly impacts the science proposed here, the primary science goals are benefited by increasing total observing baseline (assuming time observing the *Kepler* field is held fixed), since the information content of TTV signals grows rapidly with time. These details should be taken into consideration when optimizing the scientific strength of *Kepler-II*.

With our nominal target list of ~150 targets, it is likely that *Kepler-II* would be continuously observing (although with frequent changes in target selection). During this time, *Kepler-II* focuses on specific times and subarrays of high interest, e.g. transits of important planets. Our proposal can then be flexibly co-scheduled with other observations of the *Kepler* field. Indeed, a TTV-focused *Kepler-II* mission maximizes the science obtained with minimal resources due to engineering or other constraints.

## Conclusion

The *Kepler* mission has opened our eyes to a multitude of planet types and planetary architectures. To complete the study of these new worlds, we have proposed a multi-pronged observational effort called *Kepler-II*. We emphasize that the main mission has set up the opportunity to undertake dynamical studies in and around the habitable zone, as well as to find and confirm transiting planets at multiple-year periods. The campaigns of various target subsets outlined above would turn the transit method into a more powerful technique for observationally constraining planetary systems; it would allow us to measure the masses of planets for which radii have been measured; and it would measure resonant properties of pairs of planets to probe the physics of their formation and evolution.